\documentstyle[aps,preprint,epsf,rotate]{revtex}
\tightenlines

\def\d{{\rm d}}

\newcommand\as{{\alpha_s}}
\newcommand\NLO{next-to-leading order }
\newcommand\qb{{\bar q}}
\newcommand\Qb{{\bar Q}}
\newcommand{\Nc}{{N_C}}
\newcommand{\NA}{{N_A}}
\newcommand{\Nf}{{N_{\rm f}}}
\newcommand{\Nlg}{{N_{\rm lg}}}
\newcommand\gt{{\tilde g}}
\newcommand\B{{\cal B}}
\newcommand\C{{\cal C}}
\newcommand\Tr{{\rm Tr}}
\newcommand{\beq}{\begin{equation}}
\newcommand{\eeq}{\end{equation}}
\newcommand{\beqn}{\begin{eqnarray}}
\newcommand{\eeqn}{\end{eqnarray}}
\newcommand\nn{\nonumber}
\newcommand\ycut{y_{\rm cut}}

\begin{document}

\preprint{hep-ph/9708343}
\title{Excluding light gluinos using four-jet LEP events: a
next-to-leading order result}
\author{Zolt\'an Nagy$^a$ and Zolt\'an Tr\'ocs\'anyi$^{b,a}$}
\address{$^a$Department of Theoretical Physics, KLTE,
H-4010 Debrecen P.O.Box 5, Hungary
\\
$^b$Institute of Nuclear Research of the Hungarian Academy of Sciences,
H-4001 Debrecen P.O.Box 51, Hungary}
\date{\today}
\maketitle

\begin{abstract}
Based upon a next-to-leading order perturbative calculation of the
four-jet production rate in electron-positron annihilation and 
assuming 8\,\% for the theoretical error emerging from hadronization
effects in the $0.002\le \ycut \le 0.004$ range for the Durham clustering
algorithm, we exclude the existence of the light gluinos at the 95\,\%
confidence level.
\end{abstract}

%\narrowtext      

%\section{Introduction}

In recent years several interesting phenomena were observed that could be
naturally explained by the existence of light (with mass of the order of
few GeV) supersymmetric partner of the gluon --- the gluino
\cite{Farrar,Stirling}.  If light gluino exists, it should be
produced at LEP. Gluinos do not couple directly to quarks, but a gluon
can decay into a gluino pair. Thus the numerical value of four-jet
observables should be different if gluinos exist than predicted by pure
Quantum Chromodynamics (QCD). In particular, the measurement of the
eigenvalues of the Casimir operators of the underlying Lie group, the
color charges is based upon measuring four-jet observables, and this
should also be influenced by the existence of light gluinos. This fact
have been utilized by all LEP experiments for setting exclusion limits
for the existence of light gluinos \cite{L3DELPHI,OPAL,ALEPH}.
However, these analyses were based upon leading order calculation of
the four-jet observables and as a consequence the exclusion limits are
debated \cite{OPAL,ALEPH,Fodor,GFarrar}.

Recently Dixon and Signer \cite{DS4jets} and
also ourselves \cite{NT4jets} have completed the
calculation of four-jet cross sections in electron-positron
annihilation at the \NLO accuracy.  In this letter we extend the second
of these works to including the effect of light gluinos. We present the
cross sections in terms of group independent kinematical functions
multiplying the eigenvalues of the Casimir operators of the underlying
Lie group. The knowledge of the O($\alpha_s^3$) corrections to the
group independent kinematical functions facilitates the simultaneous
precision meaurement of the strong coupling and the color charge
factors using the four-jet LEP or SLC data assuming any number of light
gluino flavors. Thus in a full analysis the fit of the theoretical
prediction to the data can reveal the number of light gluino flavors
favoured by data. In this letter we do not perform such a full
analysis. Instead, we point out that in the small $\ycut$ region
($\ycut \le 0.004$) already the four-jet rate predicted for the
light-gluino scenario significantly differs from the recent ALEPH data
\cite{ALEPHR4}, while the QCD prediction is in good agreement with the
experimental results.

The main ingredients of the calculation are the four-parton \NLO and
five-parton Born level squared matrix elements. The tree level
amplitudes for the processes $e^+e^-\to \qb q ggg$ and
$e^+e^-\to \qb q \Qb Q g$ have been known for a long time
\cite{a5parton}. The new techniques developed by Bern, Dixon and Kosower
 in the calculation of one-loop multiparton amplitudes \cite{BDKannals}
made possible the derivation of explicit analytic expressions for the
helicity amplitudes of the $e^+e^-\to Z^0,\,\gamma^*\to$ 4 partons
processes \cite{BDKW}. Also, Campbell, Glover and Miller made
FORTRAN programs for calculating higher order corrections to the 
$e^+e^-\to \gamma^*\to$ 4 partons processes publicly available
\cite{CGM}. In all of these calculations, and also in our work, the
quark and lepton masses are set to zero.

We use the matrix elements of refs.~\cite{BDKW} for the loop
corrections of the QCD subprocesses. As pointed out in
refs.~\cite{Fodor,DS4jets}, the naive $\Nf=5 \to \Nf=8$ shift for
taking into account the effect of the light gluinos is not valid
exactly at the \NLO accuracy. Analyzing the color structure of the
diagrams involving gluinos and using the results of refs.~\cite{BDKW}
we have derived the necessary loop corrections for the process
$e^+e^-\to Z^0,\,\gamma^*\to \qb q \gt \gt$ and the modifications
induced by the presence of the gluinos on the four-parton one-loop QCD
amplitudes \cite{NTloopamps}. We have also derived the
$e^+e^-\to \qb q \gt \gt g$ tree-level amplitudes.

It is well known that the \NLO correction is a sum of two integrals ---
the real and virtual corrections --- that are separately divergent (in
the infrared) in $d=4$ dimensions. For infrared safe observables, for
instance the four-jet rates used in this work, their sum is finite. In
order to obtain this finite correction, we use a slightly modified
version of the dipole method of Catani and Seymour \cite{CSdipole} that
exposes the cancellation of the infrared singularities directly at the
integrand level. The formal result of this cancellation is that
the \NLO correction is a sum of two finite integrals,
\beq                         
\label{sNLO}
\sigma^{\rm NLO} 
= \int_5 \d\sigma_5^{\rm NLO} + \int_4 \d\sigma_4^{\rm NLO}\:,
\eeq
where the first term is an integral over the available five-parton
phase space (as defined by the jet observable) and the second one is
an integral over the four-parton phase space.

%\section{Results}

Once the phase space integrations in eq.~(\ref{sNLO}) are carried out,
the \NLO differential cross section for the four-jet observable $O_4$
takes the general form
\beqn
&&\frac{1}{\sigma_0}\frac{\d \sigma}{\d O_4}(O_4)
= \left(\frac{\as(\mu) C_F}{2\pi}\right)^2 \B_{O_4}(O_4)
\\ \nn &&\qquad
+ \left(\frac{\as(\mu) C_F}{2\pi}\right)^3
\left[\B_{O_4}(O_4) \frac{\beta_0}{C_F}
 \ln\frac{\mu^2}{s} + \C_{O_4}(O_4)\right]\:.
\eeqn
In this equation $\sigma_0$ denotes the Born cross section for the
process $e^+e^-\to \qb q$, $s$ is the total c.m.\ energy squared, $\mu$
is the renormalization scale, while $\B_{O_4}$ and $\C_{O_4}$ are
scale independent functions, $\B_{O_4}$ is the Born approximation
and $\C_{O_4}$ is the radiative correction. We use the two-loop
expression for the running coupling,
\beq
\as(\mu) = \frac{\as(M_Z)}{w(\mu)}
\left(
1-\frac{\beta_1}{\beta_0}\frac{\as(M_Z)}{2\pi}\frac{\ln(w(\mu))}{w(\mu)}
\right)\:,
\eeq
with
\beq
w(\mu) = 1 - \beta_0
\frac{\as(M_Z)}{2\pi}\ln\left(\frac{M_Z}{\mu}\right)\:,
\eeq
\beq
\beta_0 = \frac{11}{3}C_A
- \frac{4}{3} \left(T_R \Nf + \frac{1}{2} C_A \Nlg \right)\:,
\eeq
\beq
\beta_1 = \frac{17}{3}C_A^2 - 2 C_F T_R \Nf
 - \frac{10}{3} C_A T_R \Nf - \frac{8}{3} C_A^2 \Nlg \:,
\eeq
for $\Nlg$ number of light gluino flavors \cite{twoloopbeta}, and with the
normalization $T_R=1/2$ in $\Tr(T^aT^{\dag b})=T_R\delta^{ab}$. The
numerical values presented in this letter were obtained at the $Z^0$
peak with $M_Z=91.187$\,GeV, $\Gamma_Z=2.49$\,GeV, $\sin_W^2\theta=0.23$,
$\as(M_Z)=0.118$ and $\Nf=5$ light quark flavors.

In the case of jet rates it is customary to normalize the cross section
to the O$(\as)$ total cross section,
$\sigma_{\rm tot} = \sigma_0 \left(1+\as/\pi\right)$.
Then the four-jet rate can be written as
\beqn
\label{R4}
&&R_4(\ycut) =
\frac{1}{\sigma_{\rm tot}} \int_\infty^{\ycut} \d y \frac{\d \sigma_4}{\d y}(y)
\\ \nn &&
= \left(1+\frac{\as}{\pi}\right)^{-1}
\left(\frac{\as C_F}{2\pi}\right)^2
\\ \nn &&
\left\{B_4(\ycut)
+ \left(\frac{\as C_F}{2\pi}\right)
\left[B_4(\ycut) \frac{\beta_0}{C_F}
 \ln\frac{\mu^2}{s} + C_4(\ycut)\right]\right\}\:,
\eeqn
where the scale dependence in $\as$ has been suppressed.
The Born approximation and the higher order correction are linear and
quadratic forms of ratios of the color charges \cite{NTloopamps}:
\beq                                                                  
B_4 = B_0 + B_x\,x + B_y\,y \:,
\eeq
and            
\beqn
&&C_4 = C_0 +\,C_x\,x + C_y\,y + C_z\,z              
\\ \nn && \quad\;\;
+\,C_{xx}\,x^2 + C_{xy}\,x\,y + C_{yy}\,y^2 \:.
\eeqn
The $x$ and $y$ parameters are ratios of the quad\-ratic Casimirs,
$x=C_A/C_F$ and $y=T_R/C_F$, while $z$ is related to the square of a
cubic Casimir,
\beq
C_3 = \sum_{a,b,c=1}^\NA \Tr(T^a T^b T^{\dag c}) \Tr(T^{\dag c} T^b
T^a)\:,    
\eeq
via $z=\frac{C_3}{\Nc C_F^3}$.
We plot the numerical values for the correction functions $C_i$  for
the case of Durham clustering algorithm \cite{durham} in Fig.~1.
Note that the $C_z$ function is completely negligible, therefore it is
not plotted. In order to exhibit the behaviour in the small $\ycut$
region, the correction functions are plotted against $L^4$,
where $L = \ln(1/\ycut)$.
%\begin{figure}
%\epsfxsize=8.5cm \epsfbox{fig1.epsi}
%\caption{Group independent kinematical functions $C_i$ for QCD and QCD +
%light gluino theory.}
%\end{figure}

Using these group independent kinematical functions one can fit the
observed data for the best values of the color factor ratios $x$ and $y$
in both theories separately, and test how the obtained results are
compatible with the SU(3) values $x=9/4$ and $y=3/8$. In
principle, any kind of four-jet observable can be used for this purpose.
In practice, various normalized angular distributions were used
\cite{angles}. These observables have the virtue that their renormalization
scale dependence is very small already at tree level due to the
normalization, therefore they can be calculated reliably at leading order.
Indeed, this has been shown to be the case in the \NLO calculation in
ref.~\cite{Signer} and we confirm that result. We should like to
emphasize however, that the small scale dependence of the tree-level
perturbative result does not necessary mean that the results for the
exclusion of light gluinos from color charge measurement are
insensitive to the higher order corrections. Fig.~1 shows that the
$C_x$, $C_{xx}$ and $C_{xy}$ kinematical functions change
substantially, while the $C_y$ and $C_{yy}$ functions remain unchanged
with the inclusion of the additional degrees of freedom. This may modify
the result of the fit for the color charges.

We do not perform the complete color charge analysis here,
but note that this approach is slightly different from the
effective number of flavors approach, where the same best fit for both
theories is made and this unique result is compared with $x=9/4$ and
$y_{\rm eff}=3/5$ with the effective $y_{\rm eff}$ value obtained from
the $\Nf=5 \to 8$ rule. This latter method is made possible by the
simplicity of the tree-level matrix elements, but strictly speaking is
not valid in a \NLO analysis.

In this letter we simply take values in SU(3) for the color factor
ratios and using our kinematical functions we predict the numerical
values for the Born and correction functions $B_4(\ycut)$ and
$C_4(\ycut)$ in both theories. These functions are then fitted
according to the formula
\cite{geneva}
\beq
F(\ycut) = \sum_{n=0}^{N_F} k_n L^n\:,
\eeq
where $F=B_4$, or $C_4$.  The coefficients $k_n$ are tabulated in
Table~I. The fit-range is 0.002--0.1. The QCD results agree with the
rates obtained by Dixon and Signer \cite{DS4jets}.

Using the $B_4$ and $C_4$ functions, we construct $R_4(\ycut)$
according to formula (\ref{R4}).
We see that for small values of resolution parameter, $\ycut \le 0.004$,
the predictions for the QCD and QCD + light gluino case start to differ. 
For small values of $\ycut$ however, the fixed order perturbative
prediction is not reliable, because the expansion parameter $\as \ln^2
\ycut$ logarithmically enhances the higher order corrections. The usual
remedy is the all order resummation of the leading and next-to-leading
logarithmic contributions and matching the resummed and fixed order
results \cite{Catani}. We use R-matching as described in
refs.~\cite{Catani,DS4jets}.

Fig.~2 shows the effect of resummation and matching in QCD ($\Nlg=0$).
We see that the resummed and matched  prediction agrees with the ALEPH
data corrected to hadron level \cite{ALEPHR4} within hadronization
uncertainty. Encouraged by this, in the same figure we plot the fixed
order and the resummed and matched SU(3) predictions for the QCD +
light gluino scenario ($\Nlg=1$).  The bands indicate the theoretical
uncertainty of the fixed order prediction due to the variation of the
renormalization scale $x_\mu=\mu/\sqrt{s}$ between 1/2 and 2. The
data are well
%\input Fittable
%\noindent
described by the QCD prediction in the whole $\ycut$ range,
but they fall outside the QCD + light gluino band.
For values of $\ycut \le 0.004$ the theoretical
bands bifurcate making the discrepancy between the two theories significant.
%\begin{figure}
%\epsfxsize=8.5cm \epsfbox{fig2.epsi}
%\caption{Full \NLO and resummed-matched four-jet rate in QCD +
%$\Nlg = 0,1$ compared to ALEPH data.}
%\end{figure}

In order to quantify the exclusion level of the very light gluino, we
plot the scale dependence of the \NLO prediction at a fix value of
$\ycut=0.002$ in Fig.~3. Our choice for this value of the resolution
parameter comes from the observation that in Fig.~2 the intercept of
the $\Nlg=0$ theoretical curves for the O$(\as^3)$ and O$(\as^3)$+NLLA
prediction at $x_\mu=1$ is around 0.002, showing that in this region
the fixed order prediction itself can be trusted. Also around
$\ycut=0.002$ the hadronization effects are still small
\cite{WebberA,meannjets}.

From Fig.~3 we see that the QCD prediction agrees well with the
measured data, while the QCD + light gluino prediction falls
significantly below. Taking the renormalization scale dependence as the
theoretical error, the light gluino is excluded at more than $4 \sigma$,
or 99.99\,\% confidence level. The theoretical prediction for the
two theories differ at the $\simeq 85$\,\% confidence level. The
$\ycut=0.002$ value at LEP corresponds to a $k_\perp \simeq 5\,$GeV,
where the theoretical error emerging form hadronization does not exceed
8\,\% \cite{WebberA}. Therefore, with hadronization uncertainty included
the QCD + light gluino prediction is still excluded at the $2 \sigma$,
or 95\,\% confidence level. According to Fig.~2, the effect of
resummation of leading and next-to-leading $\ln(1/\ycut)$ logarithms is
to make the exclusion even more significant.
%\begin{figure}
%\epsfxsize=8.5cm \epsfbox{fig3.epsi}
%\caption{Renormalization scale dependence of the next-to\-leading order
%prediction for the
%four-jet rate at $\ycut=0.002$. The grey band indicates the value
%measured by ALEPH, corrected to hadron level.}
%\end{figure}

Finally, we would like to remark that the inclusion of the non-zero
gluino mass decreases the tree-level cross section \cite{Stirling}.
The K factor for the QCD + light gluino theory is small, $K\simeq 1.25$,
suggesting that this mass effect is bound to be robust when including the
higher orders.

%\section{Conclusion}

In this letter we presented the \NLO corrections to the gauge independent
kinamatical functions of the four-jet rate cross section in QCD and in
QCD with the extension of light gluinos for the case of the Durham
clustering algorithm. The knowledge of these corrections facilitates
the simultaneous precision meaurement of the strong coupling and the
color charge factors using the four-jet LEP or SLC data assuming any
number of light gluino flavors. Assuming standard SU(3) values for the
color factors, we presented the four-jet rates as predicted by \NLO
perturbation theory matched with all order resummed results in the 
next-to-leading logarithmic approximation. These theoretical
predictions were compared to recent data obtained by the ALEPH
collaboration. The four-jet rates in the QCD and QCD + light
gluino scenario are significantly different for small values of the
resolution parameter $\ycut$. The data favour the gluinoless world,
excluding the existence of light gluinos at the 95\,\% level. This
result was argued to be robust against hadronization and gluino mass
effects. Our results also show that a conventional color charge
measurement based on angle distributions should give the most significant
light gluino exclusion limit if the jets are defined at small $\ycut$,
($\ycut < 0.004$).

We thank L. Dixon for providing {\em Mathematica} files of the one-loop
helicity amplitudes, furthermore S. Catani and G. Dissertori for
helpful correspondence.  This research was supported in part by the EEC
Programme "Human Capital and Mobility", Network "Physics at High Energy
Colliders", contract PECO ERBCIPDCT 94 0613 as well as by the Hungarian
Scientific Research Fund grant OTKA T-016613.

\def\np#1#2#3  {Nucl.\ Phys.\ {\bf #1}, #2 (19#3)}
\def\npproc#1#2#3  {Nucl.\ Phys.\ Proc.\ Supp.\ {\bf #1}, #2 (19#3)}
\def\pl#1#2#3  {Phys.\ Lett.\ {\bf #1}, #2 (19#3)}
\def\prep#1#2#3  {Phys.\ Rep.\ {\bf #1}, #2 (19#3)}
\def\prd#1#2#3 {Phys.\ Rev.\ D {\bf #1}, #2 (19#3)}
\def\prl#1#2#3 {Phys.\ Rev.\ Lett.\ {\bf #1}, #2 (19#3)}
\def\zpc#1#2#3  {Zeit.\ Phys.\ C {\bf #1}, #2 (19#3)}
\def\cmc#1#2#3  {Comp.\ Phys.\ Comm.\ {\bf #1}, #2 (19#3)}
\def\anr#1#2#3  {Ann.\ Rev.\ Nucl.\ Part.\ Sci.\ #1 (19#3) #2}

\begin{figure}
\epsfxsize=16cm \epsfbox{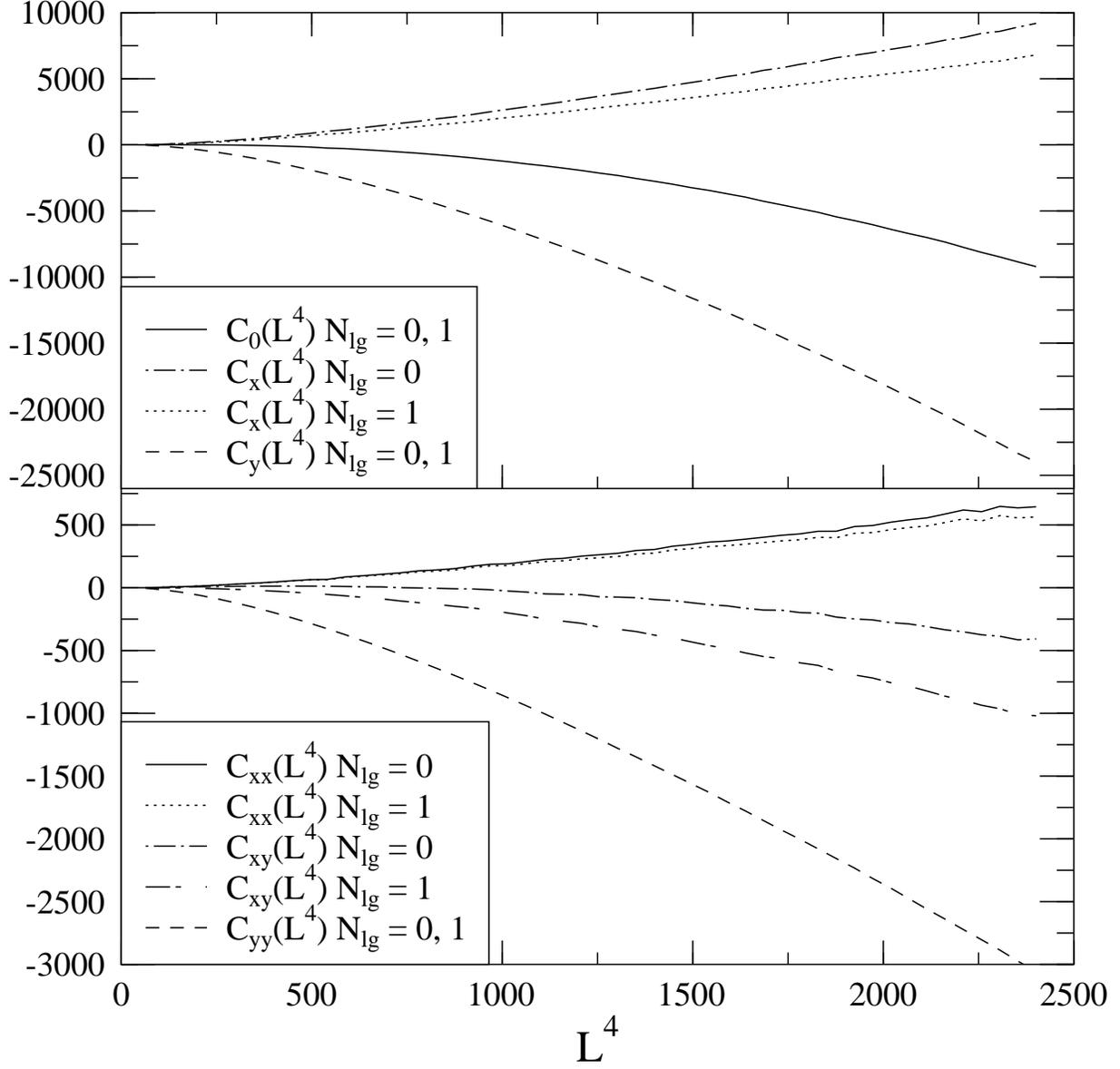}
\caption{Group independent kinematical functions $C_i$ for QCD and QCD +
light gluino theory.}
\end{figure}

\begin{table}
\caption{Coefficients of the analytic fits for the Born level and
\NLO scale independent functions $B_4$ and $C_4$. First two rows for QCD,
second two rows for QCD + light gluino theory.}
\begin{tabular}{llllllll}
   F   & $k_0$ & $k_1$ & $k_2$ & $k_3$ & $k_4$ & $k_5$ & $k_6$ \\
\tableline
 $B_4$ & -0.744& -8.224& 13.034&-6.5152& 1.0736&       &       \\
 $C_4$ & 1363  & -2742 & 2102  & -720.4& 85.2  & 5.68  &-1.0712\\
\tableline
 $B_4$ & -1.758& -7.02 & 12.663& -6.545& 1.0941&       &       \\
 $C_4$ & 528.1 &-1158.53&891.56&-250.47&-10.945& 15.821& -1.64 \\
\end{tabular}
\end{table}

\begin{figure}
\epsfxsize=16cm \epsfbox{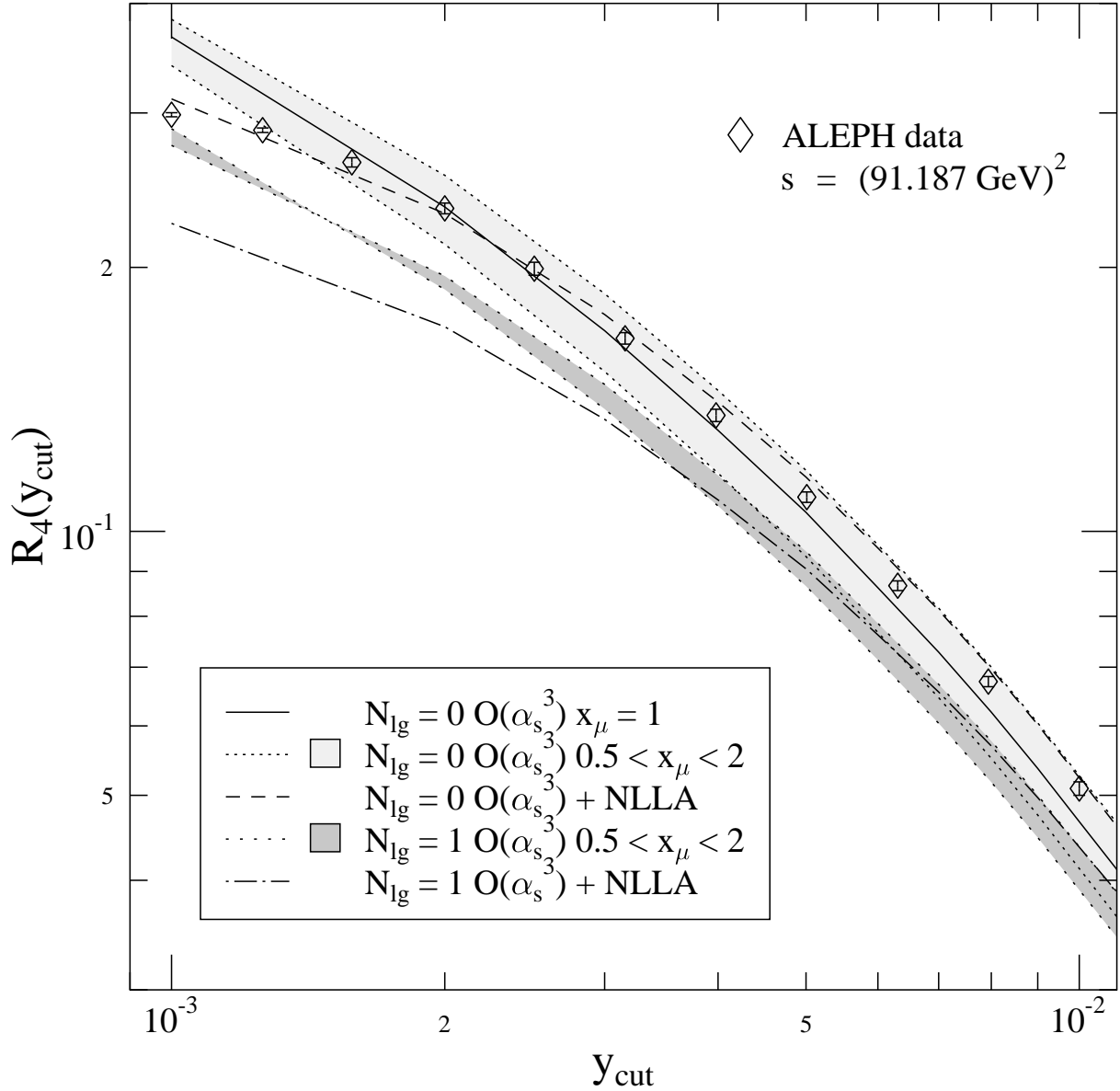}
\caption{Full \NLO and resummed-matched four-jet rate in QCD +
$\Nlg = 0,1$ compared to ALEPH data.}
\end{figure}

\begin{figure}
\epsfxsize=16cm \epsfbox{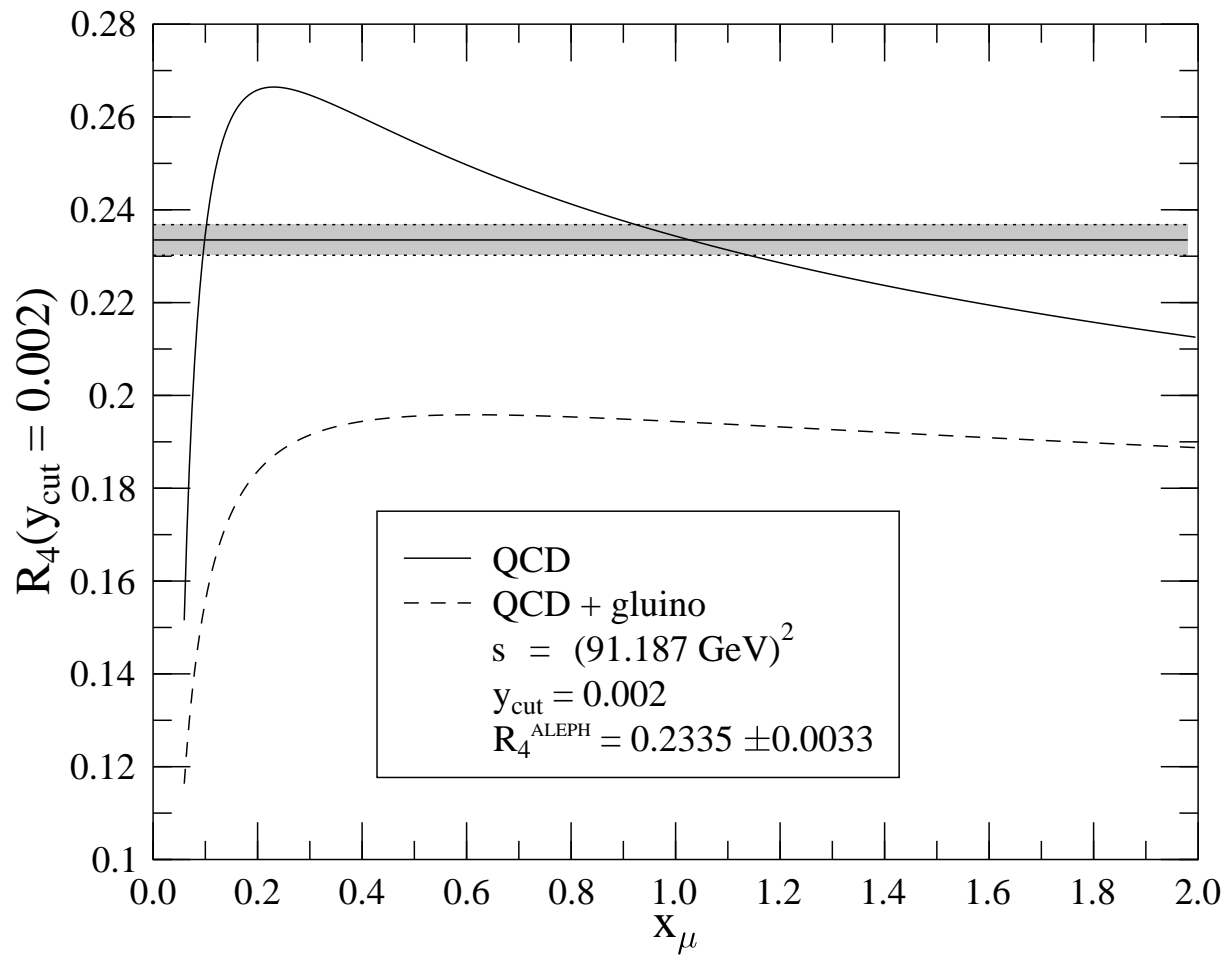}
\caption{Renormalization scale dependence of the next-to\-leading order
prediction for the
four-jet rate at $\ycut=0.002$. The grey band indicates the value
measured by ALEPH, corrected to hadron level.}
\end{figure}

\begin{thebibliography}{99}
\bibitem{Farrar}
G.R. Farrar, \pl{B265}{395}{91} ; \prd{51}{3904}{95} ;%\\
F.E. Close, G.R. Farrar and Z.P. Li, \prd{55}{5749}{97} ;%\\
G.R. Farrar, preprint hep-ph/9704309; %\\
D.J. Chung, G.R. Farrar and E.W. Kolb, preprint astro-ph/9707036;
J. Ellis, D. Nanopoulos and D. Ross, \pl{B305}{375}{93} .
\bibitem{Stirling}
R. Mu$\tilde{\rm n}$oz-Tapia and W.J. Stirling, \prd{49}{3763}{94} .
\bibitem{L3DELPHI}
B. Adeva et al, L3 Collaboration, \pl{B248}{227}{90} .
%\bibitem{DELPHI}
P. Abreu et al, DELPHI Collaboration, \zpc{59}{357}{93} .
\bibitem{OPAL}
R. Akers et al, OPAL Collaboration, \zpc{65}{367}{95} .
\bibitem{ALEPH}
R. Barate et al, ALEPH Collaboration, preprint CERN-PPE/97-002.
\bibitem{Fodor}
F. Csikor and Z. Fodor, \pl{78}{4335}{97} .
\bibitem{GFarrar}
A. de Gouvea and H. Murayama, \pl{B400}{117}{97} ;
G.R. Farrar, hep-ph/9707467.
\bibitem{DS4jets}
L. Dixon and A. Signer, \prl{78}{811}{97} ; preprint hep-ph/9706285.
\bibitem{NT4jets}
Z. Nagy and Z. Tr\'ocs\'anyi, preprint hep-ph/9707309.
\bibitem{ALEPHR4} R. Barate et al, ALEPH collaboration, preprint
CERN-PPE-96-186.
\bibitem{a5parton}
K. Hagiwara and D. Zeppenfeld, \np{B313}{560}{89} ;
F.A. Berends, W.T. Giele and H. Kuijf, \np{B321}{39}{89} ;
N.K. Falk, D. Graudenz and G. Kramer, \np{B328}{317}{89} .
\bibitem{BDKannals}
Z. Bern, L. Dixon and D. A. Kosower, \anr{46}{109}{96} .
\bibitem{BDKW}
Z. Bern, L. Dixon, D. A. Kosower and S. Wienzierl,
\np{B489}{3}{97} ;
Z. Bern, L. Dixon and D. A. Kosower, preprint hep-ph/9708239.
\bibitem{CGM} E.W.N. Glover and D.J. Miller, \pl{B396}{257}{97} ;
J.M. Campbell, E.W.N. Glover and D.J. Miller, preprint hep-ph/9706297.
\bibitem{NTloopamps} Z. Nagy and Z. Tr\'ocs\'anyi, preprint
hep-ph/9708342.
\bibitem{CSdipole}
S. Catani and M.H. Seymour, \pl{B378}{287}{96} , \np{485}{291}{97} .
\bibitem{twoloopbeta}
S. Dimopoulos, S.Raby and F. Wilczek, \prd{24}{1681}{81} ; L.E.Ibanez and
G.G. Ross, \pl{B105}{439}{81} ; W.J. Marciano and G.Senjanovic,
\prd{25}{3092}{82} ; M.B. Einhorn and D.R.T. Jones, \np{B196}{475}{82} .
\bibitem{durham} S. Catani et al, \pl{B269}{432}{91} .
\bibitem{angles} 
%\bibitem{KSW} 
J.G. K\"orner, G. Schierholz and J. Willrodt, \np{B185}{365}{81} ;
%\bibitem{NR} 
O. Nachtman and A. Reiter, \zpc{16}{45}{82} ;
%\bibitem{BZ} 
M. Bengtsson and P.M. Zerwas, \pl{B208}{306}{88} .
\bibitem{Signer} A. Signer, preprint hep-ph/9705218.
\bibitem{Catani} Catani et al, \pl{B269}{432}{91} .
\bibitem{geneva} S. Bethke et al, \np{B370}{310}{92} .
\bibitem{WebberA} B.R. Webber,
in Proceedings of the International Conference `QCD --- 20 years later',
ed.: P.M. Zerwas and H.A. Kastrup, World Scientific (1993).
\bibitem{meannjets} S. Catani et al, \np{B377}{445}{92} .
\end{thebibliography}
\end{document}